\documentclass[pdftex,12pt]{article}
\usepackage{times}
\usepackage{geometry}
\usepackage[utf8]{inputenc}
\usepackage{enumitem,amssymb}
\usepackage{ragged2e}
\usepackage[numbers,sort&compress]{natbib}
\usepackage{graphicx}
\usepackage{wrapfig}
\newlist{thematic}{itemize}{8}
\setlist[thematic]{label=$\square$}
\usepackage{pifont}
\newcommand{\cmark}{\ding{51}}%
\newcommand{\done}{\rlap{$\square$}{\raisebox{2pt}{\large\hspace{1pt}\cmark}}%
\hspace{-2.5pt}}

\newcommand{\colibri}{\textit{Colibr\`i}}

\setlist[itemize]{itemsep=2pt,wide=10pt,leftmargin=\dimexpr\labelwidth + 3\labelsep\relax,topsep=5pt}

\usepackage{titlesec}

\titleformat*{\section}{\large\bfseries}

\begin{document}
\thispagestyle{empty}
{\raggedright
\huge
Astro2020 Science White Paper \linebreak

Testing general relativity with accretion onto compact objects \linebreak
\normalsize

\noindent \textbf{Thematic Areas:} \hspace*{60pt} $\square$ Planetary Systems \hspace*{10pt} $\square$ Star and Planet Formation \hspace*{20pt}\linebreak
$\done$ Formation and Evolution of Compact Objects \hspace*{31pt} $\done$ Cosmology and Fundamental Physics \linebreak
  $\square$  Stars and Stellar Evolution \hspace*{1pt} $\square$ Resolved Stellar Populations and their Environments \hspace*{40pt} \linebreak
  $\square$    Galaxy Evolution   \hspace*{45pt} $\square$             Multi-Messenger Astronomy and Astrophysics \hspace*{65pt} \linebreak
  
\textbf{Principal Author:}

Name: Ilaria Caiazzo 
 \linebreak						
Institution: University of British Columbia 
 \linebreak
Email: ilariacaiazzo@phas.ubc.ca
 \linebreak
Phone: +16047102699 
 \linebreak
 
\textbf{Co-authors:}
Jeremy Heyl (UBC),
Adam R. Ingram (Oxford),
Tomaso Belloni (INAF Brera),
Edward Cackett (Wayne State),
Alessandra De Rosa (INAF/IAPS Roma),
Marco Feroci (INAF/IAPS Roma),
Daniel S. Swetz (NIST),
Andrea Damascelli (UBC-QMI),
Pinder Dosanjh (UBC-QMI),
Sarah Gallagher (Western/CSA),
Luigi Gallo (St Mary's),
Daryl Haggard (McGill),
Craig Heinke (Alberta),
Kelsey Hoffman (Bishop's),
Demet K{\i}rm{\i}z{\i}bayrak (UBC),
Sharon Morsink (Alberta),
Wolfgang Rau (Queens/TRIUMF),
Paul Ripoche (UBC),
Samar Safi-Harb (Manitoba),
Gregory R. Sivakoff (Alberta),
Ingrid Stairs (UBC),
Luigi Stella (INAF - Osservatorio Astronomico di Roma),
Joel N. Ullom (NIST).
}
\\
\\
\textbf{Abstract:} The X-ray emission of neutron stars and black holes presents a rich phenomenology that can lead us to a better understanding of their nature and to address more general physics questions:
\textbf{Does general relativity (GR) apply in the strong gravity regime? Is spacetime around black holes described by the Kerr metric?}
%    \item Can we better understand the physics of accretion? How do accretion disks lose angular momentum? What is the mechanism behind winds? How are jets launched?
%In particular, this white paper focuses on the following questions, and how they can be investigated with a combination of high energy and high timing resolution:
%\textbf{
%\begin{itemize}[label=$\diamond$]
%    \item Does general relativity (GR) apply in the strong gravity regime? Is space around black holes described by the Kerr metric?
%    \item Can we better understand the physics of accretion? How do accretion disks lose angular momentum? What is the mechanism behind winds? How are jets launched?
%    \item How does matter behave in extreme environments in terms of density, gravity and magnetic fields? What is the physics of ultra-dense matter? What are the masses and radii of neutron stars? 
This white paper considers how we can investigate these questions by studying reverberation mapping and quasi-periodic oscillations in accreting systems with a combination of high-spectral and high-timing resolution.
In the near future, we will be able to study compact objects in the X-rays in a new way: advancements in transition-edge sensors (TES) technology will allow for electron-volt-resolution spectroscopy combined with nanoseconds-precision timing.

\pagebreak
\pagenumbering{arabic}

\section{Introduction}
Neutron stars and black holes (\textit{compact objects}) are among the most fascinating and puzzling objects in the Universe. They uniquely provide an environment to test the laws of physics at their extremes, as density in a neutron star reaches values several times higher than nuclear density, magnetic fields are billions of times higher than the Sun's, and gravity around black holes is so strong as to trap light itself.
%This extreme nature, combined with the possibility of using them as a space laboratory to test fundamental physics, is what keeps these objects interesting to researchers and will do so in the next ten years. 
Compact objects, however, do not like to reveal their secrets all at once. Fifty years after their discovery, we still do not know what neutron stars are made of, and the question of how black holes modify space and time around them is still open.

The X-ray emission of compact objects presents a rich phenomenology that can lead us to a better understanding of their nature and to address more general physics questions:
\textbf{
\begin{itemize}[label=$\diamond$]
    \item Does general relativity (GR) apply in the strong gravity regime? Is spacetime around black holes described by the Kerr metric?
%    \item Can we better understand the physics of accretion? How do accretion disks lose angular momentum? What is the mechanism behind winds? How are jets launched?
%In particular, this white paper focuses on the following questions, and how they can be investigated with a combination of high energy and high timing resolution:
%\textbf{
%\begin{itemize}[label=$\diamond$]
%    \item Does general relativity (GR) apply in the strong gravity regime? Is space around black holes described by the Kerr metric?
%    \item Can we better understand the physics of accretion? How do accretion disks lose angular momentum? What is the mechanism behind winds? How are jets launched?
%    \item How does matter behave in extreme environments in terms of density, gravity and magnetic fields? What is the physics of ultra-dense matter? What are the masses and radii of neutron stars? 
\end{itemize}}
{\noindent The question whether GR applies to compact objects has very profound implications. Most of our fundamental theories in physics have a range of applicability and break down at some particular scale. For example, electroweak theory separates into electromagnetism and weak theory at energies lower than about 200 GeV, while strong and electroweak theories are thought to unify at very high energies. Since most of the current tests of GR are performed in the weak gravitational fields present in our solar system, the 6 to 7 orders of magnitude higher gravitational potential and $\sim$20 orders of magnitude greater curvature \cite{2015ApJ...802...63B} found around compact objects offer the tantalizing prospect of finding new gravitational physics in strong fields. }
%If there is a smoking gun for some unknown physics hidden in the strong regime, observing compact objects is our best shot at finding it.

Much progress has been made in this sense in the field of gravitational waves. However, the wavelength of gravitational waves is necessarily comparable to the size of the colliding objects, which therefore limits the scope for probing in detail the spacetime surrounding compact objects; therefore, electromagnetic signals provide a crucial complementary window into strong gravity. One of the most subtle consequences of GR is the ``no-hair'' theorem, for which black holes can be fully characterized by their mass, angular momentum and charge. Since we expect no charge on astrophysical black holes, the spacetime that surrounds a black hole can be nearly exactly described by the Kerr metric. The only way to test this theorem is to probe the spacetime very close to the hole. Fortunately, the X-ray emission of accreting black holes carries information about the inner region of the accretion disk, within a few gravitational radii ($R_g=GM/c^2$) from the hole, encoded in the fast variability of its spectrum. In particular, the emission from the accretion flow very close to accreting compact objects presents two possible high-precision diagnostics of their spacetime: reverberation mapping and quasi-periodic oscillations (QPOs).

Variability in the X-ray emission from accreting compact objects also carries information on the accretion processes themselves.
%; studying these systems can further our understanding of the physics of accretion.
Accretion disks and jets are ubiquitous in astrophysics. They are found around newborn stars, during planetary formation, in active galactic nuclei (AGNs), in which they play a key role in shaping the evolution of galaxies. However, the mechanisms for angular momentum transport in the disk and for jet formation close to the central object are poorly understood. As for any type of physics, studying accretion at its most extreme actualization, the inner accretion disk close to compact objects, provides the best opportunity for breakthroughs in the understanding of the phenomenon as a whole. For example, the high magnetic field expected to be present in the disk of accreting stellar mass black holes, and even more so around accreting neutron stars, could make it easier to highlight the role of magnetic fields in generating the viscosity needed for accretion to occur. Or the role of the black hole spin in powering winds and jets can be better understood once we measure the spins of many black holes.

In this context, recent developments in TES technology will open a new window on X-ray variability. Taking advantage of the physics of superconductors, TES-based detectors can measure the arrival time of a photon and its energy with unprecedented precision. To put things in context, a TES-based telescope with the current technology could achieve an energy resolution more than 40 times better than that of the best X-ray spectrometer in space right now (sub-eV at $\sim 1.5$ keV and about 2-3 eV in the 5-10 keV range \cite{doi:10.1063/1.4984065} compared to the 130 eV at 6 keV of \textit{XMM-Newton}) while achieving the timing resolution of the best X-ray timing telescope currently in orbit (better than 300 ns \cite{doi:10.1063/1.4962636} compared to the 100 ns of \textit{NICER}); and there is still space for improvement.  A recently proposed telescope concept, \colibri, will team TES as X-ray detectors with non-focusing optics, in order to obtain high throughput and high time and energy resolution. \colibri\  will be dedicated to the study of compact objects, and will have, among its objectives, the goal of studying the reverberation signals and quasi-periodic oscillations in accreting neutron stars and black holes, in order to study the physics of the accretion processes and the effects of strong gravity on the accretion flow.
% And there is still space for improvement. A team of astronomers have recently proposed a new design for a TES-based telescope, named \textit{Colibr\`i}, to the Canadian Space Agency. The proposal was selected for an 18-months study that started on September 4, 2018, and the team has already reached out to include scientists from Europe and the US. The goal of \textit{Colibr\`i} is to study compact objects with a combination of high precision timing and spectroscopy, thanks to the TES-based detectors, and at the same time to take advantage of the high photon rate from bright sources, thanks to the high throughput ($\sim$100kHz count-rates) achievable by teaming TES detectors with non-focusing optics.

%The spectral resolution and the large collecting area at 6 keV available with \colibri, will allow us to measure the FWHM and bulk velocity flow of multiple emitting regions around supermassive black holes. %(e.g. BLR, NLR, obscuring torus). 
%Using the neutral Fe Ka emission line, we will be able to disentangle the different components of the line and, in turn, to map the complex and rich environment of the AGN. \colibri\ also represents a powerful tool to investigate the astrophysical BHs and its connection with the galaxy (the so -called feedback Fabian, 2012, ARAA, 50, 455) through the measure of fast AGN-driven winds and outflows.

\section{Reverberation mapping}
In black-hole X-ray binaries and AGNs, accretion to the central black hole takes place via a geometrically thin, optically thick accretion disk, which emits thermally in the soft X-rays for black hole binaries and in the optical and UV bands for AGNs \cite{1973A&A....24..337S,1973blho.conf..343N}. The photons emitted by the disk are thought to be Compton up-scattered in an optically thin corona, which produces a power-law spectrum in the hard X-rays~\cite{1975ApJ...195L.101T,1979Natur.279..506S}. Some of the up-scattered photons in the corona are reflected back into the line of sight by the disk. This reflection emission presents particular features, that include an iron K$\alpha$ fluorescence line at 6.4 keV and a reflection hump that peaks at $\sim$30~keV
%, formed via inelastic scattering from free electrons
\cite{2005MNRAS.358..211R,2013ApJ...768..146G}. Gravitational redshifts from the black hole and relativistic motion of the orbiting plasma in the inner disk distort the spectrum, providing insight on the dynamics of the accretion disk \cite{1989MNRAS.238..729F}. The coronal emission shows rapid aperiodic variability, on timescales of milliseconds for stellar mass black holes and of minutes for AGNs. Neutron stars also accrete via a disk, but in addition, the material accreted onto the surface causes repeating thermonuclear reactions that are observed as bright bursts of X-ray emission, with timescales that go from 100 ms to several hours \cite{1976ApJ...206L.135B,1976ApJ...205L.127G,1976Natur.263..101W,1977xbco.conf..127M,2003A&A...405.1033C,2006ApJ...646..429C,2008ApJS..179..360G,2008int..workE..32K}. Similar to the coronal emission for black holes, this X-ray-burst emission can be reflected by the accretion disk, and indeed reflection spectra have been observed \cite{2004ApJ...602L.105B,2010ApJ...720..205C,2017ApJ...836..111K,2018ApJ...855L...4K}.

\iffalse
\begin{wrapfigure}[19]{l}{0.5\textwidth}
  \centering
  \vspace{-0.3cm}
     \includegraphics[width=0.5\textwidth]{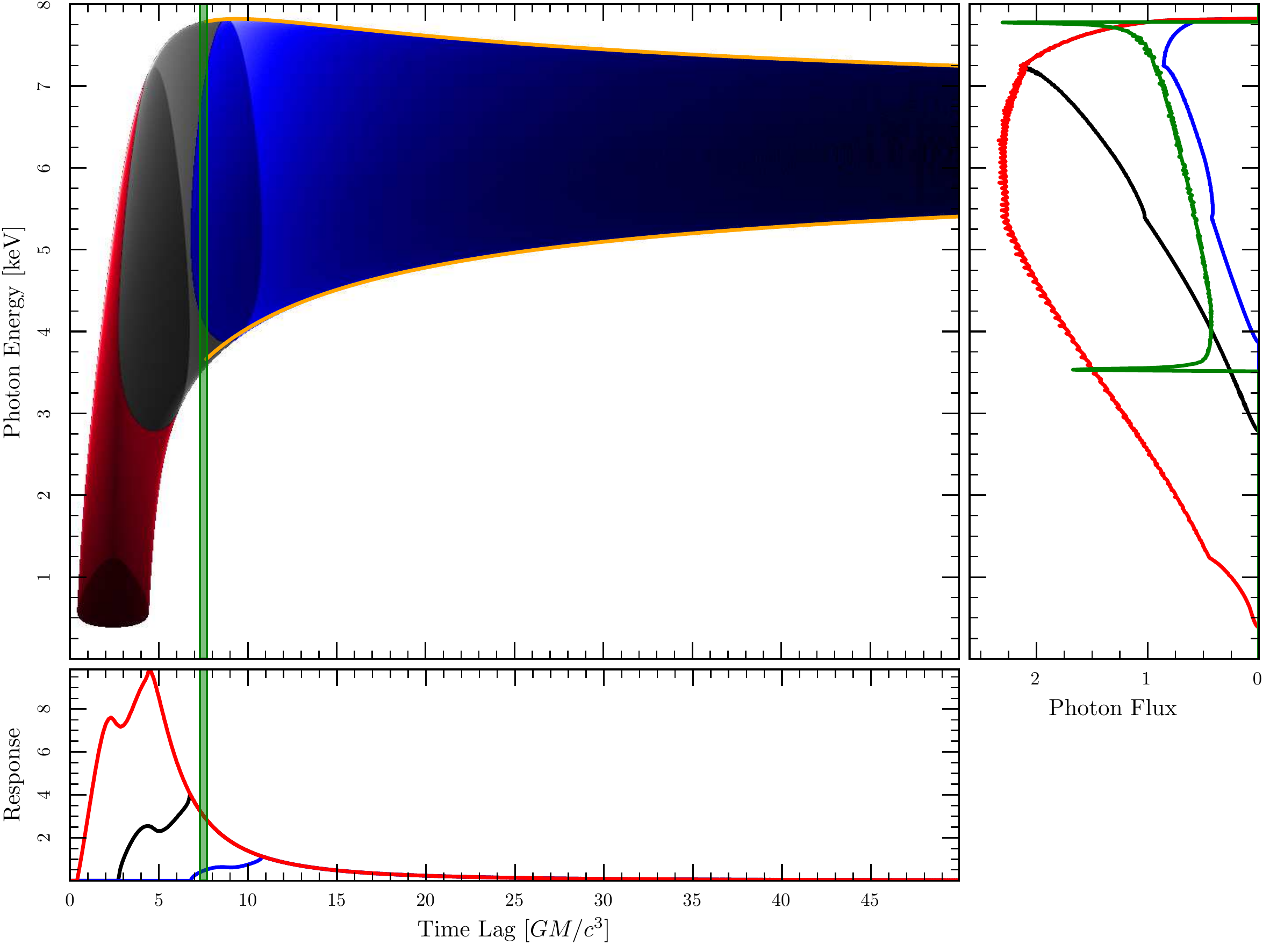}
    \vspace{-1cm}
    \caption{{\fontsize{10}{12}\selectfont Reverberation response spectrum for a black hole with spin parameter $a/M=0.998$ and an accretion disk ending at the ISCO (red), $4 GM/c^2$ (black) and $8 GM/c^2$ (blue).  The corona is modelled as a point source along the spin axis at $z=2 GM/c^2$ (lamppost), and the observer is located at an inclination of 86~degrees. See also \cite{2018MNRAS.475.4027M}. The caustics form at the edge of the response function as can be seen in the reverberation spectrum measured a particular time lag.}}
    \label{fig:rev}
\end{wrapfigure}
\else
\begin{figure}
  \centering
     \includegraphics[width=0.9\textwidth,clip,trim=0 0.5in 0 0]{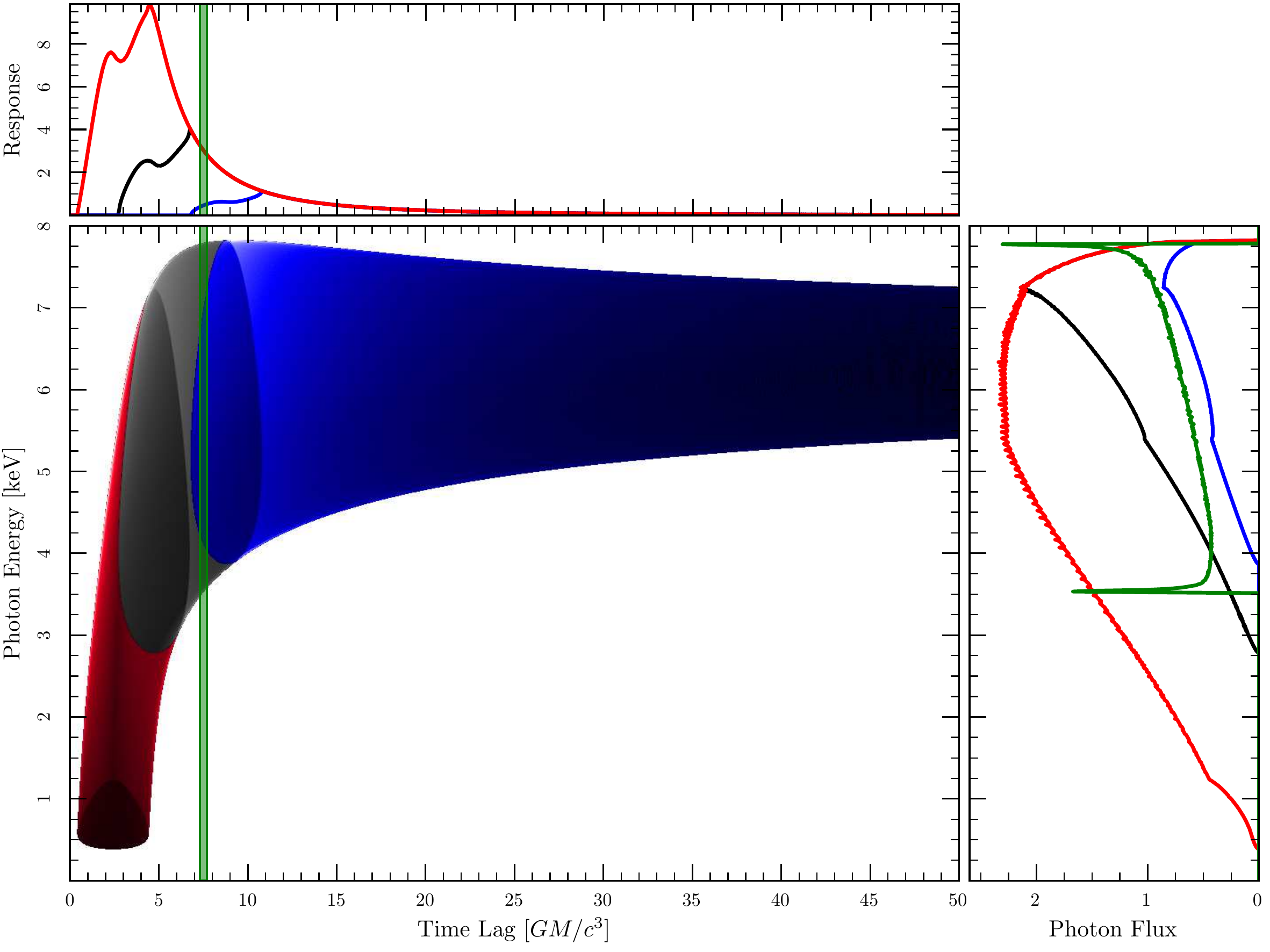}
     \includegraphics[width=0.9\textwidth]{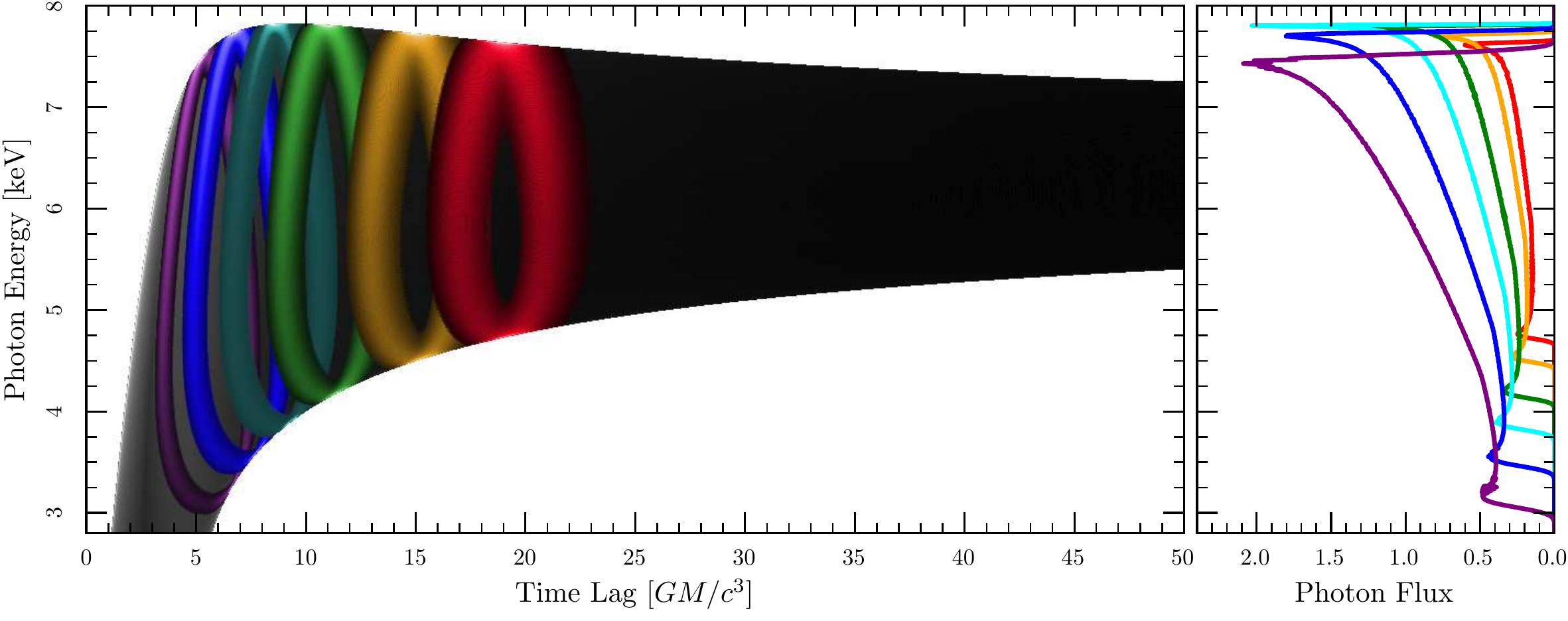}
    \vspace{-0.5cm}
    \caption{Energy- and time-dependent reflected emission resulting from a $\delta$-function (the iron line) in the driving continuum for a black hole with spin parameter $a/M=0.998$. The corona is modelled as a point source along the spin axis at $z=2 GM/c^2$ (lamppost), and the observer is located at an inclination of $86^\circ$. See also \cite{2018MNRAS.475.4027M}. {\bf Upper part:}  {\it Central panel}. Reverberation response spectrum for the accretion disk ending at the ISCO (red), $4 GM/c^2$ (black) and $8 GM/c^2$ (blue).  {\it Upper panel}. Energy integrated flux. Photons  of  different  energies,  coming from different radii on the disk, also arrive at different times. {\it Right panel}. Time integrated spectrum. The caustics form at the edge of the response function as can be seen in the reverberation spectrum measured a particular time lag (green).  {\bf Lower part:}  {\it Central panel}. The colored regions depict the reverberation signal from parts of the disk where a QPO of $Q\sim 10$ and the nodal precession frequency of 1, 2, 5, 10, 20 and 40~Hz would exist for a 10~M$_\odot$ black hole, assuming the RPM model. {\it Right panel}. The caustics can be seen in the reflection spectrum measured for a particular QPO.}
    \label{fig:rev}
\end{figure}
\fi

In both cases, variability of the illuminating signal provides a way to map the inner regions of the accretion disk, since fluctuations in the continuum emission are reflected in the reverberation spectrum with a light-crossing time delay\cite{2014A&ARv..22...72U}. Time delays can be of the order of a few hundreds of microseconds for Galactic binaries, and are much longer for AGN (scaling as $\propto M$). Also, as different parts of the accretion disk will be illuminated in subsequent times, photons of different energy will present different time delays, reflecting the characteristic Doppler shift of the reflection region. Such reverberation lags have been detected in several AGNs with \textit{XMM-Newton} and \textit{NuSTAR} \cite{2012MNRAS.422..129Z,2016MNRAS.462..511K}, and very recently in an X-ray binary with \textit{NICER} \cite{2019Natur.565..198K}.

Observations so far are limited by either the low sensitivity or low timing resolution of current X-ray telescopes. High-resolution spectral fitting of the X-ray emission, especially of the Fe-line profile, provides information on the strong-field gravity effects on the orbiting plasma and its dynamics, from which radii can be inferred in units of the gravitational radius, as well as the spin of the black hole and the inclination angle of the system. The possibility of performing reverberation mapping, which yields distances in absolute units given by the light travel time, simultaneously to spectral fitting would therefore provide a test of the Kerr metric itself, as well as a measurement of the mass of the compact object \cite{2016PhRvD..93d4020H,2016PhRvD..93l3008J}. Although the total spectral features of reverberation are relativistically broadened in general, high energy and timing resolution combined would allow slicing the emission in both the spectral and time domain, which could reveal sharp features, broadened not by the bulk motion of the disk material but by thermal and turbulent motion within the disk with $v/c \sim 10^{-3}$ or smaller as shown by the green curves in Fig.~1. 

\section{Quasi periodic oscillations}
QPOs are nearly periodic fluctuations commonly observed in the X-ray light curve from the inner regions of accreting compact objects \cite{1991ApJ...374..741M,1997ApJ...489..272T,2005Ap&SS.300..149V,2006csxs.book...39V}. The first to be discovered were the low frequency QPOs (LFQPOs), with frequencies $\sim 0.1-30$ Hz.  \textit{RXTE} revealed higher frequency features: high frequency QPOs (HFQPOs) in the range $\sim 40-450$ Hz from black hole systems \cite{1997ApJ...482..993M}, and kHz QPOs in the range $\sim 300-1200$ Hz in neutron star systems \cite{2000ARA&A..38..717V,2005Ap&SS.300..149V}. The origin of QPOs is still debated, but their frequencies are commensurate with those of orbital and epicyclic motions in the Kerr metric close to the compact object, and thus constraining the QPO mechanism would provide a new way to measure properties of the inner accretion flow and the effects of strong gravity.

Current models for LFQPOs find their origin either in some instability in the accretion flow, or in a geometric oscillation \cite{1993ApJ...417..671C,1998ApJ...492L..59S,1999ApJ...524L..63S,1999A&A...349.1003T,2001ApJ...559L..25W,2006ApJ...642..420S,2009MNRAS.397L.101I,2010MNRAS.404..738C} such as, most notably, Lense-Thirring precession \cite{1998ApJ...492L..59S,1999ApJ...524L..63S,2006ApJ...642..420S,2001ApJ...559L..25W,2009MNRAS.397L.101I}. This is a nodal precession of orbits inclined to the equatorial plane caused by a spinning compact object dragging the surrounding spacetime around with it (the frame dragging effect). The origin of HFQPOs and kHz is more obscure, with the proposed models including Doppler modulation of orbiting hotspots in the inner disk, oscillation modes of a pressure-supported torus, nonlinear resonances, gravity and pressure modes in the accretion disk \cite[][and references therein]{2013IAUS..290...57L,2014SSRv..183...43B} and, for the case of neutron stars, beating with the neutron star spin frequency \cite{2006AdSpR..38.2675V}.

It is clear that new and better observations are needed to understand these phenomena and ultimately exploit them as diagnostics. LFQPOs are normally detected with high significance using current instruments, thanks to their high amplitudes. This enables studies of the QPO phase dependence of the spectral shape (i.e. QPO tomography) \cite{2016MNRAS.461.1967I,2017MNRAS.464.2979I}. Instruments such as \textit{Colibr\`i} will revolutionize such studies by providing vastly better spectral resolution and dramatically higher count rates, particularly considering that instruments such as \textit{XMM-Newton} (and \textit{ATHENA} in the future) are limited by photon pile-up and therefore cannot be used to observe the brightest sources. %This will enable greater precision without introducing any of the systematics potentially associated with the long exposure times currently required.
% I wanted to tone-down the argument here because NICER actually solves the pile-up / deadtime problem very nicely, and we are working towards some very interesting results with it.
Furthermore, the unprecedented count rates will for the first time enable similar tomographic analyses with HF and kHz QPOs, providing a qualitatively new way of testing models. HFQPOs are much fainter than LF, and this explain the scarcity of current detections. The high sensitivity of \textit{Colibr\`i} will enable the detection of HFQPOs in more systems and it will test the presence of the even weaker signals predicted by some of the current theoretical models. The observed HFQPOs with RXTE show a similar set of frequencies from all the sources. 
%$\nu_\theta$ less than 50~Hz, and $\nu_\mathrm{per}$ and $\nu_K$ less than 400~Hz. 
This may be a selection effect from the \textit{RXTE} band, or it could mean that HFQPOs are excited only at specific frequencies. A higher sensitivity and timing resolution could bring to the detection of additional signals or to a null detection at higher frequencies, allowing to discern between the proposed models. 
%Perhaps most important in the study of QPOs is detecting large number of photons and high time resolution; nevertheless, combining this with %high energy resolution could allow a new type of diagnostic on the origin of QPOs: if the modulation is coming from a small area in the disk, %slicing the emission in energy would provide information on where the perturbation is happening.

\iffalse
\begin{wrapfigure}[24]{l}{0.6\textwidth}
  \centering
  \vspace{-0.5cm}
    \includegraphics[width=0.5\textwidth]{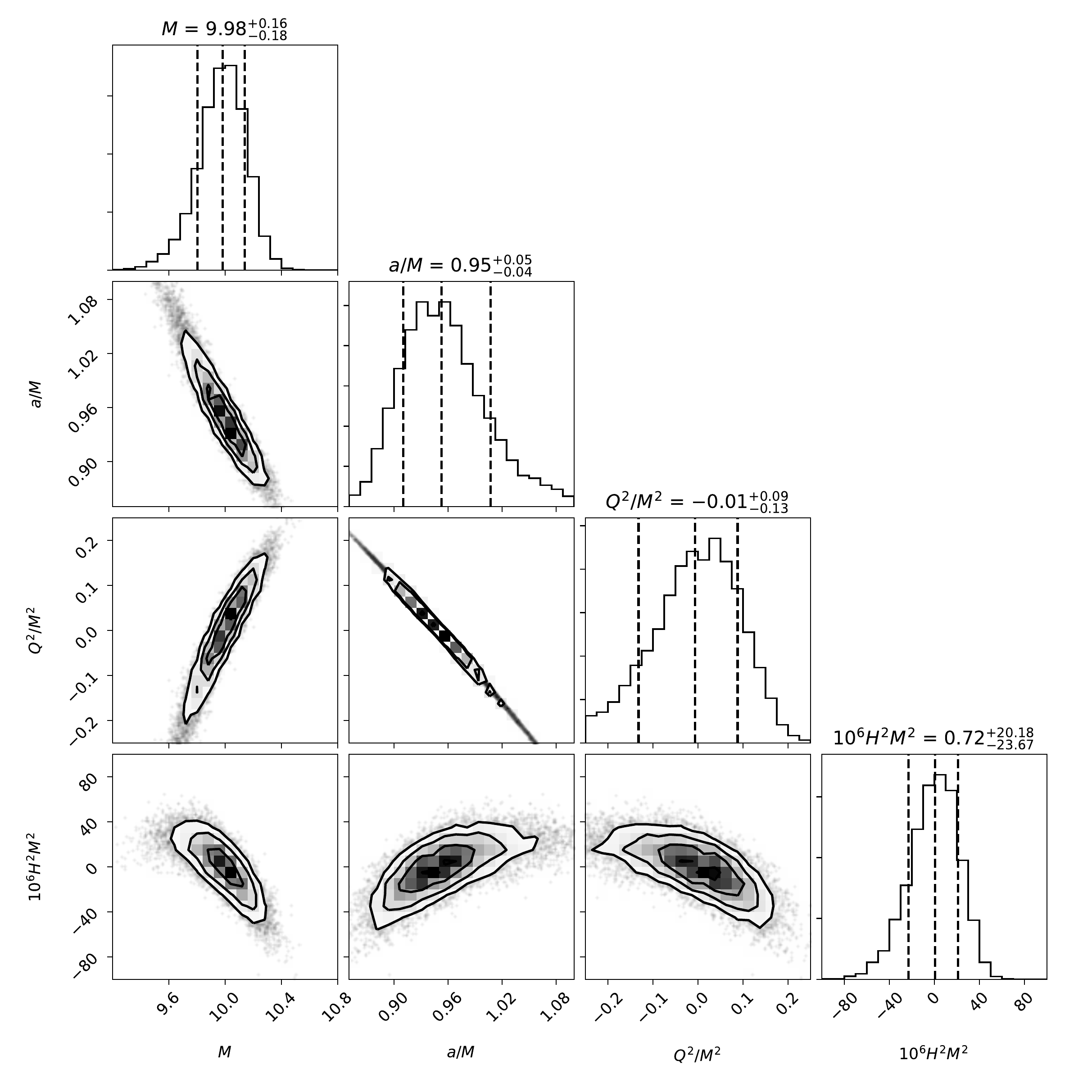}
    \vspace{-1cm}
    \caption{{\fontsize{10}{12}\selectfont Simulation of a HFQPOs for \colibri, assuming the RPM model for a black hole with spin parameter $a=0.95$ M= 10 M$_\odot$ assuming a measurement error on the frequencies of 5~Hz. The input model is the Kerr metric; the simulated frequencies are used to constrain deviation from Kerr metric by adding a cosmological constant or a charge to the black hole.}}
    \label{fig:qpos}
\end{wrapfigure}
\else
\begin{figure}[tb]
\setlength{\unitlength}{5.08cm}
\centering
\begin{picture}(3.8,2.2)
  \put(1,0.0){{\includegraphics[width=0.7\textwidth]{allparam.pdf}}}
 % \put(0.8,0.6){{\includegraphics[width=0.8\textwidth]{KN-NUT.pdf}}}
 \put(2.13,1.13){{\includegraphics[width=0.4\textwidth]{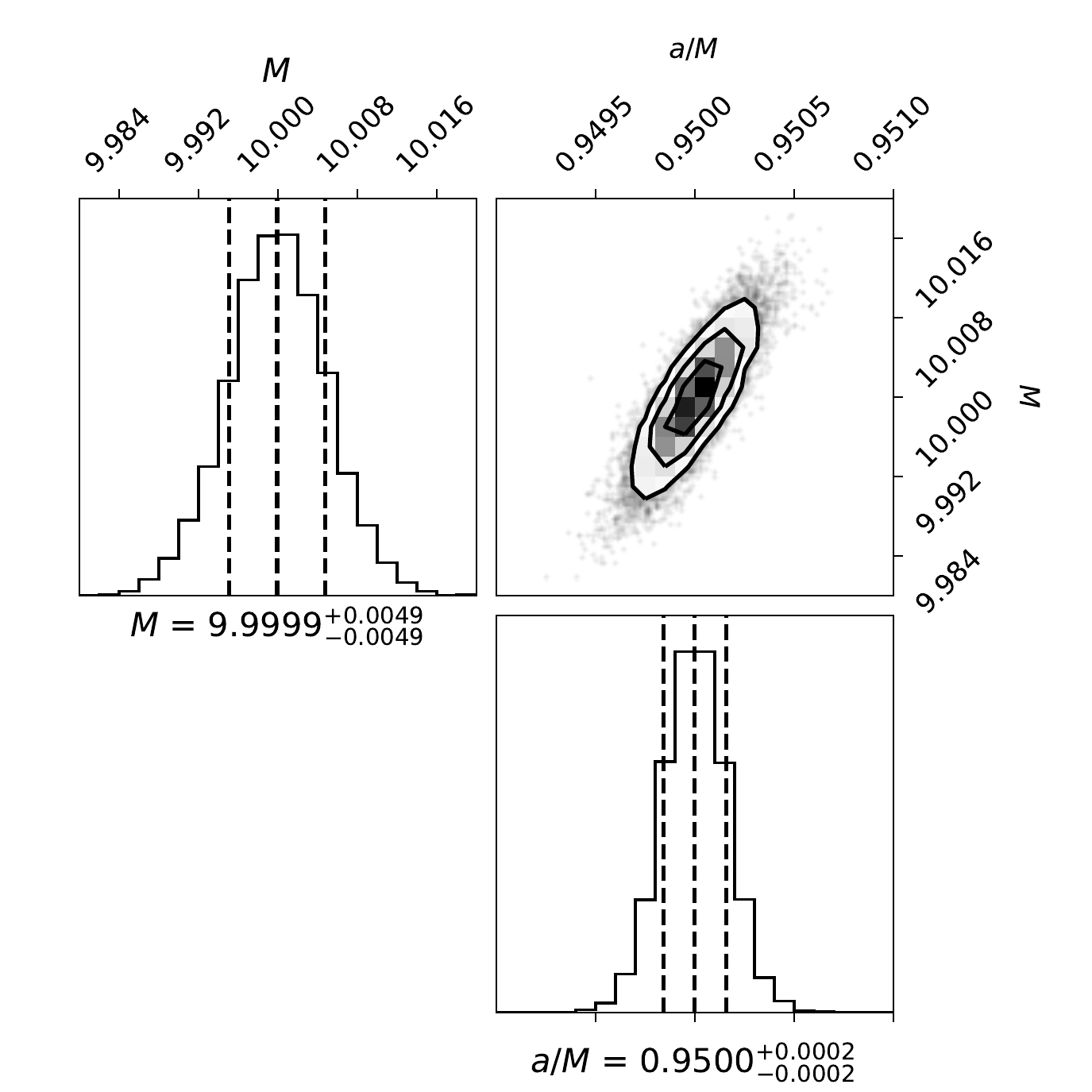}}}
  \put(-0.1,0.0){{\includegraphics[width=0.35\textwidth]{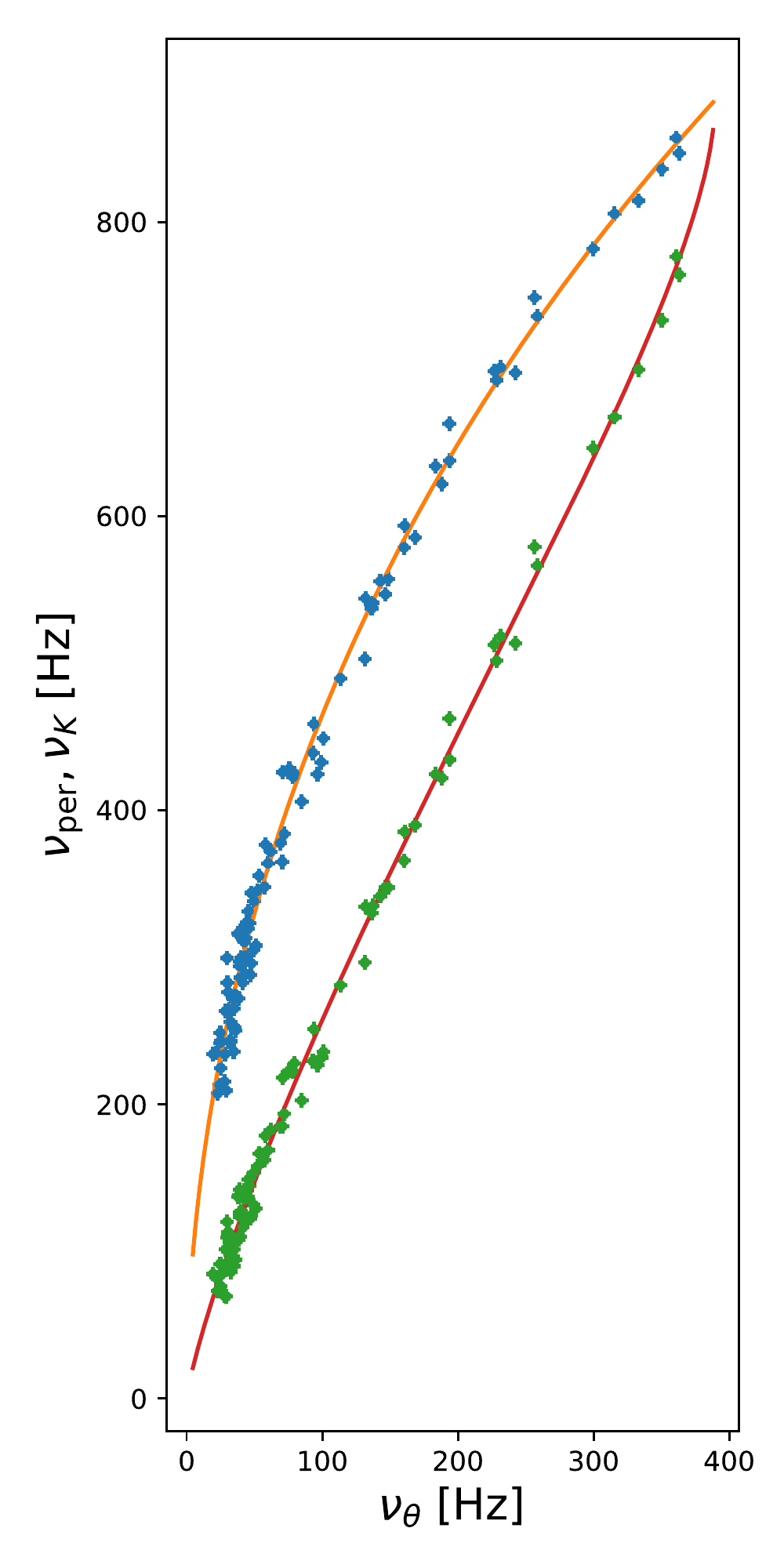}}}
  \end{picture}
    \vspace{-1cm}
    \caption{ Simulation of HFQPOs for \colibri, assuming the RPM model for a black hole with spin parameter $a=0.95$, $M=10~ \mathrm{M}_\odot$ assuming a measurement error on the frequencies of 5~Hz. The blue points depict the orbital frequency ($\nu_K$) and the green show the frequency of periastron advance ($\nu_\mathrm{per}$), both as a function of the node precession frequency ($\nu_\theta$). The input model is the Kerr metric; the simulated frequencies are used to constrain deviation from Kerr metric by adding a cosmological constant and electric charge to the black hole.}
    \label{fig:qpos}
\end{figure}
\fi
As an example of the power of QPOs as diagnostic, Fig.~\ref{fig:qpos} shows the results of a simulation for a HFQPO on a 10 M$_\odot$ BH rotating at 95\% of the critical spin performed for \textit{Colibr\`i}, where the relativistic precession model (RPM) for HFQPOs was assumed. In the RPM, four signals are expected in the power spectrum: Lense-Thirring precession, radial epicyclic motion, periastron precession and orbital motion, and the relation between the frequencies is uniquely determined by the Kerr metric. A drift in the frequencies is observed, correlated to a variability in the flux, that is interpreted in the RPM as a response to a change in the inner radius of the disk. Observations of the four signals and of how they drift, not only would shed light on the origin of the phenomenon, it would constrain the mass and spin of the hole with great accuracy, as shown in Fig.~\ref{fig:qpos}. Furthermore, since mass and spin are the only parameters important in describing the spacetime around a black hole in the Kerr metric, these observations could provide constraints on deviations from it. In Fig.~\ref{fig:qpos} a simple test is shown: adding a charge or a cosmological constant, both expected to be negligible in astrophysical black holes, changes the metric from Kerr; finding a value different from zero would hint to a deviation from the Kerr metric and from GR.

\pagebreak
\bibliographystyle{jer}
\bibliography{CSA}

\begin{thebibliography}{48}
\providecommand{\natexlab}[1]{#1}
\providecommand{\url}[1]{\texttt{#1}}
\expandafter\ifx\csname urlstyle\endcsname\relax
  \providecommand{\doi}[1]{doi: #1}\else
  \providecommand{\doi}{doi: \begingroup \urlstyle{rm}\Url}\fi

\bibitem[{Baker} et~al.(2015){Baker}, {Psaltis}, and
  {Skordis}]{2015ApJ...802...63B}
T.~{Baker}, D.~{Psaltis}, and C.~{Skordis}.
\newblock \emph{\apj}, 802:\penalty0 63, 2015.

\bibitem[Morgan et~al.(2017)Morgan, Pappas, Bennett, Gard, Hays-Wehle, Hilton,
  Reintsema, Schmidt, Ullom, and Swetz]{doi:10.1063/1.4984065}
K.~M. Morgan, C.~G. Pappas, D.~A. Bennett, J.~D. Gard, J.~P. Hays-Wehle, G.~C.
  Hilton, C.~D. Reintsema, D.~R. Schmidt, J.~N. Ullom, and D.~S. Swetz.
\newblock \emph{Applied Physics Letters}, 110\penalty0 (21):\penalty0 212602,
  2017.

\bibitem[Morgan et~al.(2016)Morgan, Alpert, Bennett, Denison, Doriese, Fowler,
  Gard, Hilton, Irwin, Joe, O'Neil, Reintsema, Schmidt, Ullom, and
  Swetz]{doi:10.1063/1.4962636}
K.~M. Morgan, B.~K. Alpert, D.~A. Bennett, E.~V. Denison, W.~B. Doriese, J.~W.
  Fowler, J.~D. Gard, G.~C. Hilton, K.~D. Irwin, Y.~I. Joe, G.~C. O'Neil, C.~D.
  Reintsema, D.~R. Schmidt, J.~N. Ullom, and D.~S. Swetz.
\newblock \emph{Applied Physics Letters}, 109\penalty0 (11):\penalty0 112604,
  2016.

\bibitem[{Shakura} and {Sunyaev}(1973)]{1973A&A....24..337S}
N.~I. {Shakura} and R.~A. {Sunyaev}.
\newblock \emph{\aap}, 24:\penalty0 337--355, 1973.

\bibitem[{Novikov} and {Thorne}(1973)]{1973blho.conf..343N}
I.~D. {Novikov} and K.~S. {Thorne}.
\newblock In C.~{Dewitt} and B.~S. {Dewitt}, editors, \emph{Black Holes (Les
  Astres Occlus)}, pages 343--450, 1973.

\bibitem[{Thorne} and {Price}(1975)]{1975ApJ...195L.101T}
K.~S. {Thorne} and R.~H. {Price}.
\newblock \emph{\apjl}, 195:\penalty0 L101--L105, 1975.

\bibitem[{Sunyaev} and {Truemper}(1979)]{1979Natur.279..506S}
R.~A. {Sunyaev} and J.~{Truemper}.
\newblock \emph{\nat}, 279:\penalty0 506--508, 1979.

\bibitem[{Ross} and {Fabian}(2005)]{2005MNRAS.358..211R}
R.~R. {Ross} and A.~C. {Fabian}.
\newblock \emph{\mnras}, 358:\penalty0 211--216, 2005.

\bibitem[{Garc{\'{\i}}a} et~al.(2013){Garc{\'{\i}}a}, {Dauser}, {Reynolds},
  {Kallman}, {McClintock}, {Wilms}, and {Eikmann}]{2013ApJ...768..146G}
J.~{Garc{\'{\i}}a}, T.~{Dauser}, C.~S. {Reynolds}, T.~R. {Kallman}, J.~E.
  {McClintock}, J.~{Wilms}, and W.~{Eikmann}.
\newblock \emph{\apj}, 768:\penalty0 146, 2013.

\bibitem[{Fabian} et~al.(1989){Fabian}, {Rees}, {Stella}, and
  {White}]{1989MNRAS.238..729F}
A.~C. {Fabian}, M.~J. {Rees}, L.~{Stella}, and N.~E. {White}.
\newblock \emph{\mnras}, 238:\penalty0 729--736, 1989.

\bibitem[{Belian} et~al.(1976){Belian}, {Conner}, and
  {Evans}]{1976ApJ...206L.135B}
R.~D. {Belian}, J.~P. {Conner}, and W.~D. {Evans}.
\newblock \emph{\apjl}, 206:\penalty0 L135--L138, 1976.

\bibitem[{Grindlay} et~al.(1976){Grindlay}, {Gursky}, {Schnopper},
  {Parsignault}, {Heise}, {Brinkman}, and {Schrijver}]{1976ApJ...205L.127G}
J.~{Grindlay}, H.~{Gursky}, H.~{Schnopper}, D.~R. {Parsignault}, J.~{Heise},
  A.~C. {Brinkman}, and J.~{Schrijver}.
\newblock \emph{\apjl}, 205:\penalty0 L127--L130, 1976.

\bibitem[{Woosley} and {Taam}(1976)]{1976Natur.263..101W}
S.~E. {Woosley} and R.~E. {Taam}.
\newblock \emph{\nat}, 263:\penalty0 101--103, 1976.

\bibitem[{Maraschi} and {Cavaliere}(1977)]{1977xbco.conf..127M}
L.~{Maraschi} and A.~{Cavaliere}.
\newblock In K.~A. {van der Hucht}, editor, \emph{X-ray Binaries and Compact
  Objects}, pages 127--128, 1977.

\bibitem[{Cornelisse} et~al.(2003)]{2003A&A...405.1033C}
R.~{Cornelisse} et~al.
\newblock \emph{\aap}, 405:\penalty0 1033--1042, 2003.

\bibitem[{Cumming} et~al.(2006){Cumming}, {Macbeth}, {in 't Zand}, and
  {Page}]{2006ApJ...646..429C}
A.~{Cumming}, J.~{Macbeth}, J.~J.~M. {in 't Zand}, and D.~{Page}.
\newblock \emph{\apj}, 646:\penalty0 429--451, 2006.

\bibitem[{Galloway} et~al.(2008){Galloway}, {Muno}, {Hartman}, {Psaltis}, and
  {Chakrabarty}]{2008ApJS..179..360G}
D.~K. {Galloway}, M.~P. {Muno}, J.~M. {Hartman}, D.~{Psaltis}, and
  D.~{Chakrabarty}.
\newblock \emph{\apjs}, 179:\penalty0 360-422, 2008.

\bibitem[{Keek} and {in't Zand}(2008)]{2008int..workE..32K}
L.~{Keek} and J.~J.~M. {in't Zand}.
\newblock In \emph{Proceedings of the 7th INTEGRAL Workshop}, page~32, 2008.

\bibitem[{Ballantyne} and {Strohmayer}(2004)]{2004ApJ...602L.105B}
D.~R. {Ballantyne} and T.~E. {Strohmayer}.
\newblock \emph{\apjl}, 602:\penalty0 L105--L108, 2004.

\bibitem[{Cackett} et~al.(2010){Cackett}, {Miller}, {Ballantyne}, {Barret},
  {Bhattacharyya}, {Boutelier}, {Miller}, {Strohmayer}, and
  {Wijnands}]{2010ApJ...720..205C}
E.~M. {Cackett}, J.~M. {Miller}, D.~R. {Ballantyne}, D.~{Barret},
  S.~{Bhattacharyya}, M.~{Boutelier}, M.~C. {Miller}, T.~E. {Strohmayer}, and
  R.~{Wijnands}.
\newblock \emph{\apj}, 720:\penalty0 205--225, 2010.

\bibitem[{Keek} et~al.(2017){Keek}, {Iwakiri}, {Serino}, {Ballantyne}, {in't
  Zand}, and {Strohmayer}]{2017ApJ...836..111K}
L.~{Keek}, W.~{Iwakiri}, M.~{Serino}, D.~R. {Ballantyne}, J.~J.~M. {in't Zand},
  and T.~E. {Strohmayer}.
\newblock \emph{\apj}, 836:\penalty0 111, 2017.

\bibitem[{Keek} et~al.(2018){Keek}, {Arzoumanian}, {Bult}, {Cackett},
  {Chakrabarty}, {Chenevez}, {Fabian}, {Gendreau}, {Guillot}, {G{\"u}ver},
  {Homan}, {Jaisawal}, {Lamb}, {Ludlam}, {Mahmoodifar}, {Markwardt}, {Miller},
  {Prigozhin}, {Soong}, {Strohmayer}, and {Wolff}]{2018ApJ...855L...4K}
L.~{Keek}, Z.~{Arzoumanian}, P.~{Bult}, E.~M. {Cackett}, D.~{Chakrabarty},
  J.~{Chenevez}, A.~C. {Fabian}, K.~C. {Gendreau}, S.~{Guillot},
  T.~{G{\"u}ver}, J.~{Homan}, G.~K. {Jaisawal}, F.~K. {Lamb}, R.~M. {Ludlam},
  S.~{Mahmoodifar}, C.~B. {Markwardt}, J.~M. {Miller}, G.~{Prigozhin},
  Y.~{Soong}, T.~E. {Strohmayer}, and M.~T. {Wolff}.
\newblock \emph{\apjl}, 855:\penalty0 L4, 2018.

\bibitem[{Mastroserio} et~al.(2018){Mastroserio}, {Ingram}, and {van der
  Klis}]{2018MNRAS.475.4027M}
G.~{Mastroserio}, A.~{Ingram}, and M.~{van der Klis}.
\newblock \emph{\mnras}, 475:\penalty0 4027--4042, 2018.

\bibitem[{Uttley} et~al.(2014){Uttley}, {Cackett}, {Fabian}, {Kara}, and
  {Wilkins}]{2014A&ARv..22...72U}
P.~{Uttley}, E.~M. {Cackett}, A.~C. {Fabian}, E.~{Kara}, and D.~R. {Wilkins}.
\newblock \emph{\aapr}, 22:\penalty0 72, 2014.

\bibitem[{Zoghbi} et~al.(2012){Zoghbi}, {Fabian}, {Reynolds}, and
  {Cackett}]{2012MNRAS.422..129Z}
A.~{Zoghbi}, A.~C. {Fabian}, C.~S. {Reynolds}, and E.~M. {Cackett}.
\newblock \emph{\mnras}, 422:\penalty0 129--134, 2012.

\bibitem[{Kara} et~al.(2016){Kara}, {Alston}, {Fabian}, {Cackett}, {Uttley},
  {Reynolds}, and {Zoghbi}]{2016MNRAS.462..511K}
E.~{Kara}, W.~N. {Alston}, A.~C. {Fabian}, E.~M. {Cackett}, P.~{Uttley}, C.~S.
  {Reynolds}, and A.~{Zoghbi}.
\newblock \emph{\mnras}, 462:\penalty0 511--531, 2016.

\bibitem[{Kara} et~al.(2019)]{2019Natur.565..198K}
E.~{Kara} et~al.
\newblock \emph{\nat}, 565:\penalty0 198--201, 2019.

\bibitem[{Hoormann} et~al.(2016){Hoormann}, {Beheshtipour}, and
  {Krawczynski}]{2016PhRvD..93d4020H}
J.~K. {Hoormann}, B.~{Beheshtipour}, and H.~{Krawczynski}.
\newblock \emph{\prd}, 93\penalty0 (4):\penalty0 044020, 2016.

\bibitem[{Jiang} et~al.(2016){Jiang}, {Bambi}, and
  {Steiner}]{2016PhRvD..93l3008J}
J.~{Jiang}, C.~{Bambi}, and J.~F. {Steiner}.
\newblock \emph{\prd}, 93\penalty0 (12):\penalty0 123008, 2016.

\bibitem[{Miyamoto} and {Kitamoto}(1991)]{1991ApJ...374..741M}
S.~{Miyamoto} and S.~{Kitamoto}.
\newblock \emph{\apj}, 374:\penalty0 741--743, 1991.

\bibitem[{Takizawa} et~al.(1997)]{1997ApJ...489..272T}
M.~{Takizawa} et~al.
\newblock \emph{\apj}, 489:\penalty0 272--283, 1997.

\bibitem[{van der Klis}(2005)]{2005Ap&SS.300..149V}
M.~{van der Klis}.
\newblock \emph{\apss}, 300:\penalty0 149--157, 2005.

\bibitem[{van der Klis}(2006)]{2006csxs.book...39V}
M.~{van der Klis}.
\newblock \emph{{Rapid X-ray Variability}}, pages 39--112.
\newblock 2006.

\bibitem[{Morgan} et~al.(1997){Morgan}, {Remillard}, and
  {Greiner}]{1997ApJ...482..993M}
E.~H. {Morgan}, R.~A. {Remillard}, and J.~{Greiner}.
\newblock \emph{\apj}, 482:\penalty0 993--1010, 1997.

\bibitem[{van der Klis}(2000)]{2000ARA&A..38..717V}
M.~{van der Klis}.
\newblock \emph{\araa}, 38:\penalty0 717--760, 2000.

\bibitem[{Chakrabarti} and {Molteni}(1993)]{1993ApJ...417..671C}
S.~K. {Chakrabarti} and D.~{Molteni}.
\newblock \emph{\apj}, 417:\penalty0 671, 1993.

\bibitem[{Stella} and {Vietri}(1998)]{1998ApJ...492L..59S}
L.~{Stella} and M.~{Vietri}.
\newblock \emph{\apjl}, 492:\penalty0 L59--L62, 1998.

\bibitem[{Stella} et~al.(1999){Stella}, {Vietri}, and
  {Morsink}]{1999ApJ...524L..63S}
L.~{Stella}, M.~{Vietri}, and S.~M. {Morsink}.
\newblock \emph{\apjl}, 524:\penalty0 L63--L66, 1999.

\bibitem[{Tagger} and {Pellat}(1999)]{1999A&A...349.1003T}
M.~{Tagger} and R.~{Pellat}.
\newblock \emph{\aap}, 349:\penalty0 1003--1016, 1999.

\bibitem[{Wagoner} et~al.(2001){Wagoner}, {Silbergleit}, and
  {Ortega-Rodr{\'{\i}}guez}]{2001ApJ...559L..25W}
R.~V. {Wagoner}, A.~S. {Silbergleit}, and M.~{Ortega-Rodr{\'{\i}}guez}.
\newblock \emph{\apjl}, 559:\penalty0 L25--L28, 2001.

\bibitem[{Schnittman} et~al.(2006){Schnittman}, {Homan}, and
  {Miller}]{2006ApJ...642..420S}
J.~D. {Schnittman}, J.~{Homan}, and J.~M. {Miller}.
\newblock \emph{\apj}, 642:\penalty0 420--426, 2006.

\bibitem[{Ingram} et~al.(2009){Ingram}, {Done}, and
  {Fragile}]{2009MNRAS.397L.101I}
A.~{Ingram}, C.~{Done}, and P.~C. {Fragile}.
\newblock \emph{\mnras}, 397:\penalty0 L101--L105, 2009.

\bibitem[{Cabanac} et~al.(2010){Cabanac}, {Henri}, {Petrucci}, {Malzac},
  {Ferreira}, and {Belloni}]{2010MNRAS.404..738C}
C.~{Cabanac}, G.~{Henri}, P.-O. {Petrucci}, J.~{Malzac}, J.~{Ferreira}, and
  T.~M. {Belloni}.
\newblock \emph{\mnras}, 404:\penalty0 738--748, 2010.

\bibitem[{Lai} et~al.(2013)]{2013IAUS..290...57L}
D.~{Lai} et~al.
\newblock In \emph{Feeding Compact Objects: Accretion on All Scales}, volume
  290 of \emph{IAU Symposium}, pages 57--61, 2013.

\bibitem[{Belloni} and {Stella}(2014)]{2014SSRv..183...43B}
T.~M. {Belloni} and L.~{Stella}.
\newblock \emph{\ssr}, 183:\penalty0 43--60, 2014.

\bibitem[{van der Klis}(2006)]{2006AdSpR..38.2675V}
M.~{van der Klis}.
\newblock \emph{Advances in Space Research}, 38:\penalty0 2675--2679, 2006.

\bibitem[{Ingram} et~al.(2016){Ingram}, {van der Klis}, {Middleton}, {Done},
  {Altamirano}, {Heil}, {Uttley}, and {Axelsson}]{2016MNRAS.461.1967I}
A.~{Ingram}, M.~{van der Klis}, M.~{Middleton}, C.~{Done}, D.~{Altamirano},
  L.~{Heil}, P.~{Uttley}, and M.~{Axelsson}.
\newblock \emph{\mnras}, 461:\penalty0 1967--1980, 2016.

\bibitem[{Ingram} et~al.(2017){Ingram}, {van der Klis}, {Middleton},
  {Altamirano}, and {Uttley}]{2017MNRAS.464.2979I}
A.~{Ingram}, M.~{van der Klis}, M.~{Middleton}, D.~{Altamirano}, and
  P.~{Uttley}.
\newblock \emph{\mnras}, 464:\penalty0 2979--2991, 2017.

\end{thebibliography}

\end{document}